# The Reaction Mechanism of the Azide-Alkyne Huisgen Cycloaddition

Martina Danese,*[a] Marta Bon,[a] GiovanniMaria Piccini[b]† and Daniele Passerone*[a]

The azide-alkyne Huisgen cycloaddition has a key role in click chemistry and configures as powerful tool in pharmaceutical and medicinal chemistry. Although this reaction has already been largely studied, there is an ongoing debate about its mechanism. In this work we study the dynamical aspects of the process using metadynamics computer simulation. We focus on the conformational aspects that determine the course of the reaction and characterize the free energy landscape of the process. To properly capture the thermodynamics of the process we select optimal collective variables using harmonic linear discriminant analysis. The results confirm the experimental evidence qualitatively and give insight on the role of the substituents and the possible transition mechanisms.

## Introduction

In the last decades, the use of alkyne-azide Huisgen cycloaddition has widely spread in several aspects of drug discovery, medicinal chemistry and biochemistry, due to its high regioselectivity and to the different chemical and physical properties of the 1,4 and 1,5 reaction products (see Scheme 1).[1-3] In organic synthesis the Cu-catalyzed reaction is used for natural compound modifications enhancing their solubility or for the synthesis of sugars macro-cycles, peptides, and acids.[2] Moreover, Ru-catalyzed reactions are mainly used in peptidomimetics design, for permeable and stable macro-cycles formation, and in nucleic acids modification.[3]

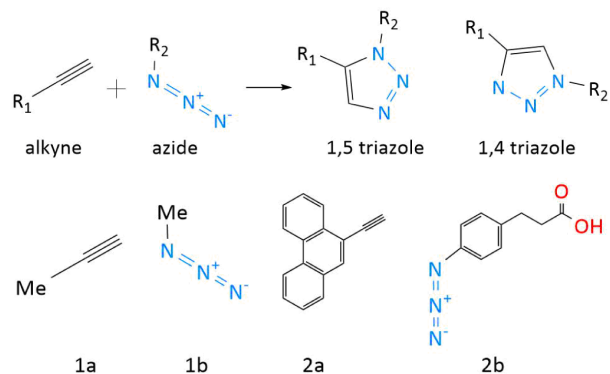

**Scheme 1.** *(a) Huisgen cycloaddition in the gas phase (b) Reactants used in this work.*

Despite the huge number of its applications, the Huisgen cycloaddition mechanism is still not fully understood even for the reaction isolated in the gas phase.[4] For 1,3-dipolar cycloadditions two possible mechanisms are hypothesized. The first suggests a concerted route[1,5] while the second implies a stepwise process involving a diradical intermediate and two distinct energy barriers.[6] Both hypotheses can explain the experimental evidence[1,6] and the currently availably computational studies are not conclusive on this point.[4] Even for the simplest 1,3-dipolar reactions, such as fulminic acid with acetylene, ab initio calculations predict a concerted process,[7] while studies based on a semi-empirical potential energy surface suggest the stepwise mechanism.[8,9] The latter outcome, however, might be affected by the neglect of closed shell diatomic overlap.[4]

This work aims at shedding light on the conformational aspects and, thus, on the thermodynamics of the Huisgen cycloaddition reaction by means of atomistic computer simulations. In computational science, a standard way of sampling the free energy landscape of a reaction is molecular dynamics (MD). Despite its simplicity and straightforward applicability, MD suffers of a severe time scale problem when the interesting metastable states, namely reactants and products, are separated by high free energy barriers. To circumvent this problem, several techniques such as well-tempered metadynamics (WT-MetaD),[10,11] have been proposed to accelerate transitions between metastable states allowing the estimation of free energy surface (FES). In WT-MetaD, a history-dependent adaptive bias is added to the underlying potential energy of the system. This bias is deposited along specific order parameters, named collective variables (CVs), able to discriminate between the metastable states and to fully characterize the thermodynamics of the process.

In a chemical reaction, reactants and products can be discriminated by their bonding topology. Therefore, a reasonable choice for the CVs could be a linear combination of a relevant subset of interatomic distances that play a substantial role in determining the breaking and formation of chemical bonds. Unfortunately, for complex reactions an *a priori* determination of the linear combination coefficients driven by chemical intuition may be rather cumbersome. Harmonic linear discriminant analysis (HLDA) provides an automated way to determine these weights and has proven effective for problems of increasing complexity.[12,13,14]

In this work, we combined HLDA in its multiclass variant[12] and WT-MetaD to study two Huisgen reactions differing in the nature of the substituents of the alkyne and azide molecules. As shown in Scheme 1, in the first reaction we used **1a** and **1b** as reactants, substituting $R_1$ and $R_2$ positions with a methyl group. The second reaction involves a phenantrene group ($R_1=C_4H_{10}$) as alkyne substituent (**2a**) and phenilpropionic acid ($R_2=C_9H_{10}O_2$) as the azide moiety (**2b**).

The comparison of the reactions **1a + 1b** and **2a + 2b** allows investigating the influence of the nature of the substituents on the stereoisomerism of the products.[15]

We studied the reaction using the semiempirical PM6 Hamiltonian[16] to determine energy and forces driving the dynamics of the system. This choice offers a reasonably good compromise between computational cost and qualitative description of the process, allowing identifying and properly characterizing the main structural motifs of the reaction.

## Methods

### A. Harmonic Linear Discriminant Analysis for Chemistry

Let us consider a generic chemical reaction involving *m* states, e.g. reactants evolving into different products. At finite temperature, a short MD run in each *i*-th local free energy minimum spans a confined area (class) of the *N*-dimensional descriptors space. Each class can be associated to an expectation value ($\mu_i$) and to a multivariate standard deviation ($\Sigma_i$) of the *N* descriptors.

Aim of the HLDA is to find the lowest dimensional projection (*W*) that best discriminates all the classes. In this way the number of descriptors is optimally reduced maintaining a faithful description of the process. Given *m* metastable states, HLDA identifies a low dimensional subspace formed by the eigenvectors of *W* associated to the *m-1* non zero eigenvalues.[12,13] These eigenvectors represent the directions along which the equilibrium metastable states distribution are best separate and, thus, can serve as optimal CVs to enhance the sampling of the reactions.

As explained in ref. 15, *W* can be found by maximizing the objective function:

$$J(W) = \frac{W^T S_B W}{W^T S_w W}, \qquad (1)$$

where $S_B$ and $S_w$ are the "between class" and "within class" matrixes, the first associated to the coordinates of the centroid of the distributions and the second to the pooled covariance of the data. For the latter, rather than taking the arithmetic average of the state covariance matrixes like in standard Fisher's LDA, a harmonic average is chosen (details and theoretical foundations of this choice can be found in refs 2 and 13):

$$S_B = \sum_i (\mu_i - \bar{\mu})(\bar{\mu} - \mu_i)^T, \quad (2)$$

$$S_w = \frac{1}{\Sigma_1 + \Sigma_2 + \cdots \Sigma_m}, \quad (3)$$

where $\bar{\mu}$ is the centroid of the whole data set, i.e. the mean of the $m$ means $\mu_i$. The projection matrix coming from Eq. (1) is diagonalized and its $m-1$ non-trivial eigenvectors will become our CVs for the optimal sampling of the FES under study.

We applied HLDA to our case of study as explained in the following.

At first, we identified the $N$ descriptors. In the case of the Huisgen cycloaddition these are all the distances ($d_{1-7}$) involved in the formation of the cycle (Figure 1)

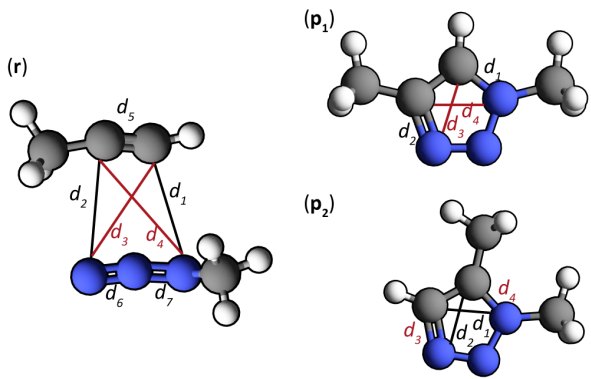

*Figure 1. 1a + 1b reactants (r) and products (p₁, p₂). $d_1$, $d_2$, $d_3$, $d_4$, $d_5$, $d_6$ and $d_7$ are the N descriptors used to find CVs with HLDA method Eqs. (1-3).*

For each metastable state in both reactions, we performed unbiased MD simulations in the NVT ensemble. These runs allow to extract $\mu_i$ and $\sigma_i$ in the 7-*th* dimensional distances space, that were further used as prescribed in Eqs. (1-3). The coefficients of the HLDA projections ($\xi_1$ and $\xi_2$) detailed in SM were proved to be converged after 140 and 200 ps, respectively for the **1a + 1b** and the **2a + 2b** system (Figure S1 of SM).

### B. Computational Details

Adopting $\xi_1$ and $\xi_2$ (outcome of the HLDA) as CVs, we performed WT-MetaD simulations in the adaptive Gaussians formulation[17] to explore the FES of the reaction. The additional parameters of the WT-MetaD calculations are reported in Table S2 of the SM. Simulations were carried out in the canonical ensemble at T=300 K, using the canonical sampling velocity rescaling thermostat,[18] and the equations of motion were implemented with a time step of 1 fs. The thermostat has been coupled very frequently (every 100 step) to mimic the effect of the environment.

In order to avoid the conformational space exploration of uninteresting areas, we applied a harmonic restraint on $d_1$, $d_2$, $d_3$ and $d_4$ for values higher than 8 Å, and another restraint to $d_5$, $d_6$ and $d_7$ to prevent products from breaking apart. In the post-processing analysis, to keep into account the effect of these external potentials together with the deposited bias, we used the reweighting procedure, as explained in ref. 15.

Free energy differences and associated statistical errors between reactants and products are obtained from the converged FES as explained in Section S2 of SM.

All calculations were carried out with the CP2K/4.0 code,[19] patched with the PLUMED2 plugin.[20]

To investigate the steric hindrance effect of the substituents, geometry optimizations were performed applying restraints on $d_1$, $d_2$, $d_3$ and $d_4$ distances, which are the atoms involved in the bond formation for the cycloaddition.

We set the harmonic constant *k* of the restraints to 250 kJ/(mol Å) and considered the geometry to be relaxed when all forces dropped below 2.40 kJ/(mol Å).

WT-MetaD simulations were carried out at the semi-empirical level using the "neglect of differential diatomic overlap" (NDDO) approximation with the PM6[16] parametrization due to its wide applicability and good overall performance.[21,22] We set the wavefunction convergence criterion to $10^{-7}$ a.u. We performed all calculations with the wavelet Poisson solver.[23]

## Results

### A. 1a + 1b reaction

**Harmonic Linear Discriminant Analysis.** As explained in the Methods section, the CVs identified with HLDA, $\xi_1$ and $\xi_2$, are linear combination of the distances $d_{1-7}$ (Table S1 of SM). Each of these corresponds to a confined area in the $d_{1-7}$ descriptor space. Figure 2 shows the capability of the two CVs to discriminate between reactants and products states. Moreover, as shown in Figure 2b for a complete discrimination both CVs are needed. As an example, projections on the $d_1$-$d_2$ (Figure 2b) and on the $d_1$-$d_3$ (Figure 2 b) spaces are reported. Figure 2b emphasizes that the projection only on the $\xi_2$ line is not enough to separate the reactants (green) and the 1,5 triazole (blue) trajectories, while the projection on $\xi_1$ does not give a separation between the two products as good as $\xi_2$.

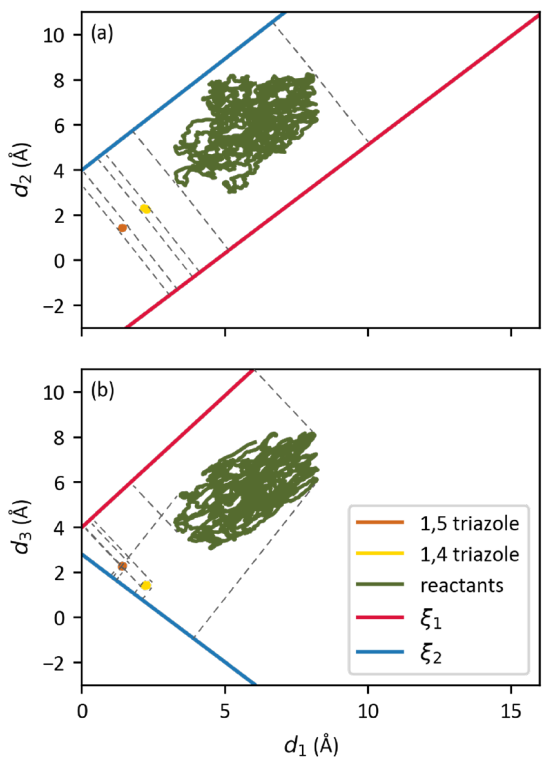

***Figure 2.*** *Examples of $d_{1-7}$ descriptors subspaces. Trajectories (green, yellow and orange), $\xi_1$, $\xi_2$ directions (blue and red) and trajectories projection lines over them are shown. (a) In $d_1$ $d_2$ space only one CV would be needed to discriminate the two products from the reactants. (b) In $d_1$-$d_3$ space both CVs are needed.*

**WT-MetaD.** To give a flavor of the WT-MetaD technique, we report in Figure 3 the configurational space exploration in the initial stages of the simulation. As shown, the different reaction states can be recognized by the associated value of the CVs and occupy limited regions on the configurational space. While the shape of the products (green and orange) is more elliptical, the area occupied by the reactants configurations (blue) is broaden, due to the highly fluxional character (and consequential freedom) of the two initial molecules. This justifies the need of overcoming the usual harmonic approximation that barely takes into account the Hessian structure around the PES minima to compute the entropic contribution to the FES.

The resulting converged FES are reported in Figures 4 and 7. In particular, from Figure 4 it is clear that the two products basins have similar depth, and quantitative analysis confirms that they are equally populated (Table 1).

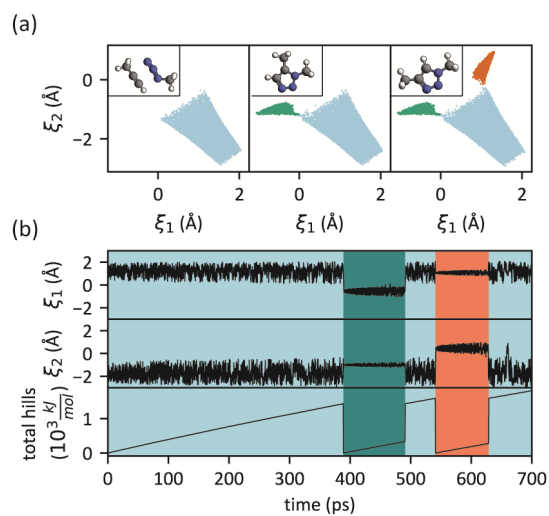

*Figure 3. (a) Configurational space exploration in WT-MetaD simulation as function of the $\xi_1$ and $\xi_2$ CVs. Representative geometries of the system at the different simulation times are reported in the upper right corner of each image. (b) Behavior of the CVs and repulsive bias added in time. Here the first 700 ps are represented for the **1a + 1b** reaction.*

In Figure 4 we observe a high free energy intermediate, **t\***, separating the two products. This configuration has only been scarcely explored during the simulation and did never lead to the formation of the reaction products.

*Table 1. Resulting basins free energy differences (ΔF) from WT-MetaD simulations and potential energy differences (ΔE) from structural optimizations. Reactants are set as reference and units are in kJ/mol.*

|   | ΔF | | ΔE | |
|---|---|---|---|---|
|   | 1,4 triazole | 1,5 triazole | 1,4 triazole | 1,5 triazole |
| **1a + 1b** | − 176 ± 2 | − 180 ± 10 | −222 | −224 |
| **2a + 2b** | −147 ± 7 | −152 ± 4 | −205 | −234 |

From the resulting FES, one can identify two possible reaction mechanisms towards the two products. The first goes *directly* from the reactants basin to the products and accordingly to the final product is labelled as $m_{1,4}^d$ and $m_{1,5}^d$. The second requires the reactants to pass through a lateral *channel* ($a_{1,4}$ and $a_{1,5}$) at the edge of the reactants basin and is referred as $m_{1,4}^c$ and $m_{1,5}^c$. In these lateral corridors the two reactants form a Van der Waals complex with the carbon chain of the alkyne parallel to the nitrogen chain of the azide, a configuration accessible from the reactants basin, by overcoming a rotational barrier of 100-150 kJ/mol. These important features of the reaction mechanism have emerged in a natural way from metadynamics, and could have not been easily captured by simpler methods, based on educated guesses (e.g. transition state theory).

Following the two possible paths ($m^d$ and $m^c$), it is trivial to identify two regions (one for each product) that correspond to intermediate configurations, possible candidates for transition states ($t_{1,4}$ and $t_{1,5}$). As shown in Figure 5, the geometries in these regions differ in the bonding topology ($N_3$-$C_4$ and a $N_1$-$C_5$) and are associated to different reweighted-probabilities[24] (see section S3 of SM for more details on the post process calculations).

In the ensemble of transition structures, we found two possible geometries leading to the formation of 1,4 triazole, the first ($t_{1,4}^{s1}$) corresponding to a *stepwise* mechanism and the second ($t_{1,4}^c$) to a *concerted* process that exhibits the highest occurrence probability. Even in the transition ensemble towards 1,5 triazole we observe two structures associated to the two possible mechanisms, but in this case the associated probabilities do not exhibit large differences as in the case of $t_{1,4}$. Note that the two structures associated to stepwise processes ($t_{1,4}^{s1}$ and $t_{1,5}^{s2}$) differ for their bonding topology. While $t_{1,4}^{s1}$ shows $N_1$-$C_5$ bond, $t_{1,5}^{s2}$ exhibits a $N_3$-$C_4$ bond, evidence that can be explained in light of steric hindrance. To graphically represent this concept (Figure 6), we performed further restrained geometry optimizations (see Section 2.2.2 for details on the calculation), and calculated the corresponding potential energy surfaces (PES). These show high energy regions separating reactants and products. The minimum energy paths of the reaction should pass through the intermediate regions $I_1$ (with geometries similar to $t_{1,4}^{s1}$) and $I_2$ for the 1,5 triazole (with geometries similar to $t_{1,5}^{s2}$).

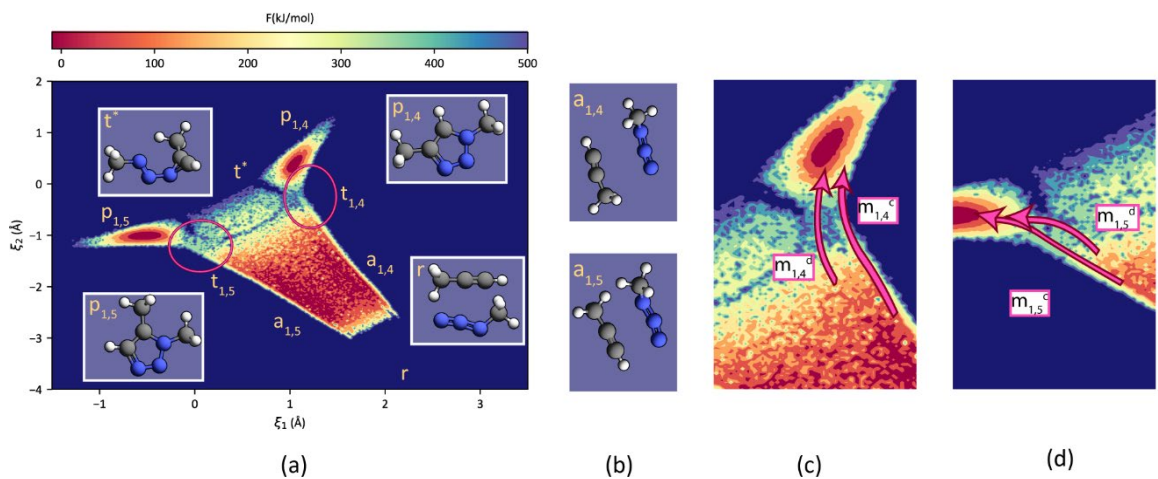

***Figure 4.*** *(a) Reweighted FES for **1a** + **1b** system. **r** and **p** indicate reactants and products, while the lateral channels are labelled as **a**. A high free energy intermediate is also identified and is called **t\***. Areas in the pink circles correspond to the transition regions. (b) $a_{1,4}$ and $a_{1,5}$ representative configurations. (c) and (d) The "direct" mechanism from reactants to products is labelled as $m^d$, while the indirect, requiring the passage through the lateral corridors **a** is named with $m^c$. The subscripts indicate if the corresponding area is involved in the reaction towards 1,4 or 1,5 triazole.*

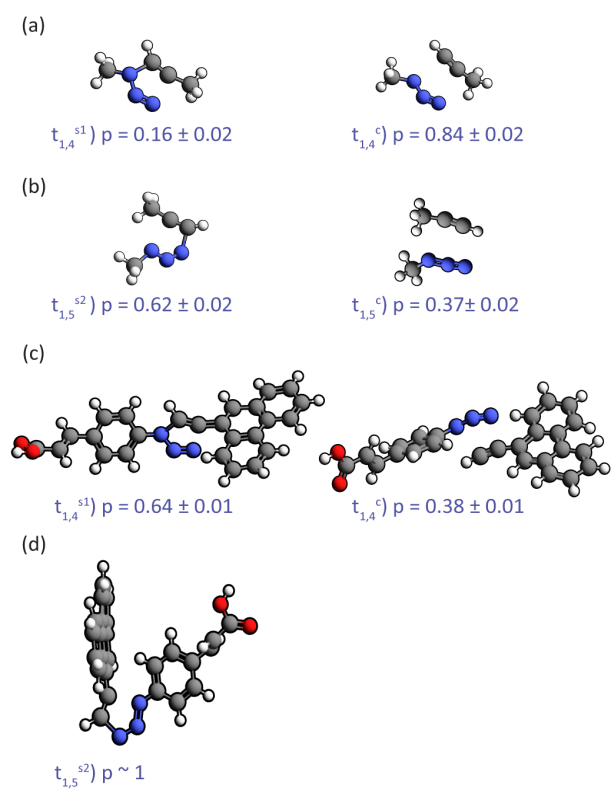

***Figure 5.*** *(a) and (b) Geometries found in $t_{1,4}$ and $t_{1,5}$ regions of Figure 4. (c) and (d) Structures associated to the $t_{1,4}$ and $t_{1,5}$ regions of Figure 7.*

We note that the area associated to $I_1$ exhibits higher values of the PES compared to the ones of $I_2$. This can be a source for the difference in the probabilities of the stepwise structures $t_{1,4}^{s1}$ and $t_{1,5}^{s2}$.

On the contrary, a direct explanation of the surprisingly high probability of the concerted configurations ($t_{1,4}^c$ and $t_{1,5}^c$) based merely on the PES analysis is not possible: indeed, Figure 6 shows no evidence of possible saddle points in the concerted region (along the diagonal). The corresponding configurations must then be favored on the FES due to thermal effects (enthalpic and entropic contributions).

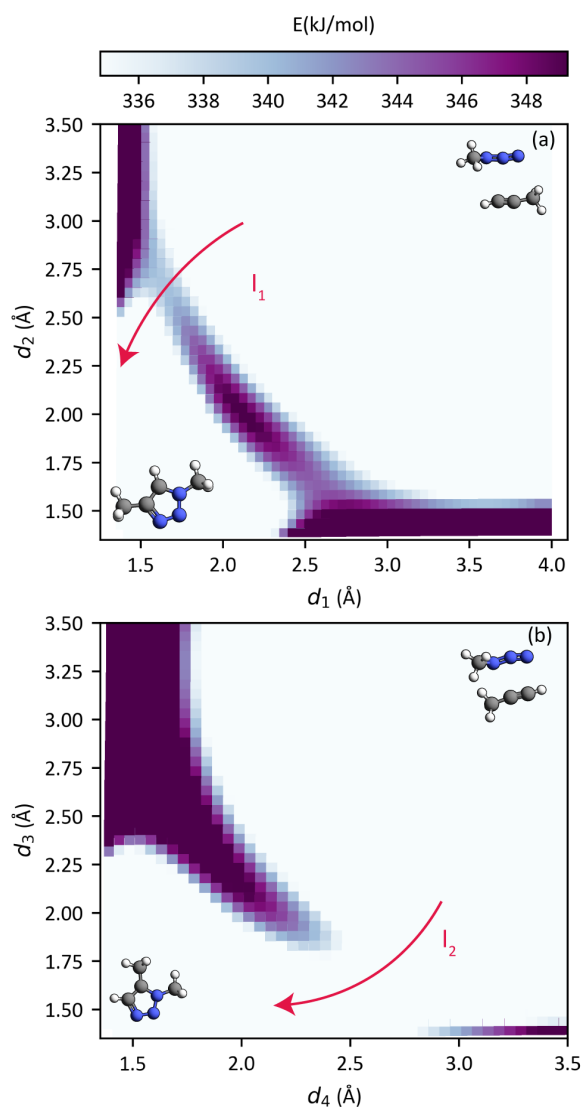

*Figure 6. Energy grids generated from geometry optimizations with restraints on $N_1$-$C_5$ and $N_3$-$C_4$ distances. The color scale indicates the potential energy corresponding to each point of the grid, and the different ranges are chosen to highlight the presence of the saddle points. The energy wall towards 1,5 triazole in (b) grid and its absence towards 1,4 triazole in (a) can be explained with the steric hindrance between the two substituents in 1,5 triazole configuration.*

### B. 2a + 2b reaction

The larger substituents in **2a** and **2b** (Scheme 1) do not bring any change in the overall shape of the FES (compared to the one associated to **1a + 1b** reaction) and on the relative positions of its basins.

It is worth observing a visible difference between the two FES, i.e. the two maps are rotated in the CVs space. This does not represent a problem, as the hypersurface identified by HLDA CVs ($\xi_1$ and $\xi_2$) is invariant under rotations.[13]

As shown in Table 1, the two products exhibit the same free energy differences compared to the reactants, similarly to what happens for the **1a + 1b** reaction. The main difference between the two systems is instead in the heat of reaction. We observe that **2a + 2b** reaction is less exothermic compared to **1a + 1b** and we attribute this effect to the increased entropy in the reactants and reduced in the products, both associated to the size of the substituents.

Concerning the structures observed in the transition areas of the FES, some differences are recovered compared to the case of **1a + 1b** reaction. We believe that this results from the size of the substituents. Their enhanced steric hindrance has the effect of reversing the probability trend for $t_{1,4}$ (thus favoring the stepwise over the still possible concerted mechanism, Figure 5c), and making the concerted mechanism not accessible for $t_{1,5}$ (Figure 5d). Nevertheless, the geometries associated to the stepwise processes show the same topology as the ones of the **1a + 1b** reaction ($t_{1,4}^{s1}$ and $t_{1,5}^{s2}$ in Figure 5c), indicating that the change of substituents does not significantly modify the PES of the reaction (Figure S4 of SM), but significantly contributes to the free energy to the finite temperature probabilities).

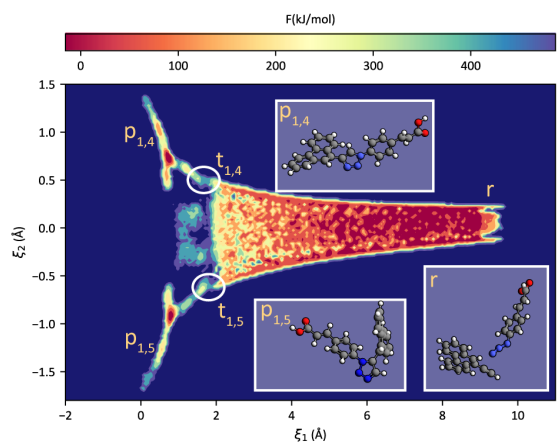

***Figure 7.*** *Reweighted FES for **2a + 2b** system. The discussion about the transition state configuration is equivalent to the **1a + 1b** and reported in the supplementary material.*

## Conclusions

In this work we used well-tempered metadynamics to study the mechanism of two Huisgen cycloadditions differing in the nature and size of the substituents. Metadynamics allows capturing on the free energy landscape of the system the salient conformational features that are essential for a realistic description of a complex chemical reaction.

The resulting free energy surfaces show for both systems that 1,4 and 1,5 triazole are equally populated within the statistical error.

The configurations visited by metadynamics in the transition regions from reactants to products can be traced back as belonging to a "stepwise" or a "concerted" mechanism.

The probability values associated to such structures indicate a preferential concerted process towards the 1,4 triazole, and a stepwise mechanism to 1,5 triazole when the substituents are methyl groups. The change to larger functional groups ($R_1=C_4H_{10}$ and $R_2=C_9H_{10}O_2$) seems to invert the concerted/stepwise preference for 1,4 triazole and to kill the concerted for 1,5 triazole.

Further investigation of the potential energy surface (PES) underlying the reaction highlighted the role of steric hindrance and underlines the need of including thermal effects for a correct description of the reaction, since no saddle points are apparent in the "concerted" region of the PES.

Does this evidence solve the conundrum about the mechanism of the Huisgen reaction, dating back to the debate between Firestone and Huisgen? Not completely: a discussion on the zero temperature transition states (individuation of saddle points on the PES) can only be conclusive at a higher level of theory (e.g., multiconfigurational quantum chemistry methods involving charge separation and radical character). Nevertheless, the precious information gained by finite temperature free energy exploration (focus of the present paper) could not have been obtained without resorting to semiempirical approximations. This choice allowed us to prove that for such fluxional systems thermal effects overwhelm potential energy effects in determining the overall reaction features.

For this reason we believe that this computational strategy will be applicable to relevant reactions of the same kind even at increased level of complexity, such as in presence of catalytic surfaces. For example, chiral intermetallic compounds showed to be enantioselective with respect to the adsorption of similar molecules as the ones described here [Prinz, J., Gröning, O., Brune, H. & Widmer, R. Highly Enantioselective Adsorption of Small Prochiral Molecules on a Chiral Intermetallic Compound. *Angew. Chem.* **127**, 3974-3978 (2015)] and are presently being experimentally investigated with respect to surface-supported Huisgen type reactions.

## Acknowledgments

We acknowledge M. Parrinello and A. C. Corminboeuf for their scientific advice. This work was supported by the NCCR MARVEL. In particular, M.D. was supported by the MARVEL INSPIRE Potentials Masters fellowship. We would like to thank also R. Widmer and S. Stolz for helpful discussions and for sharing unpublished experimental results.

# Supporting Materials

## S1 Harmonic Linear Discriminant analysis

The convergence of HLDA coefficients was tested monitoring in time the absolute value of the scalar product between the instantaneous value of each CV and its final one.

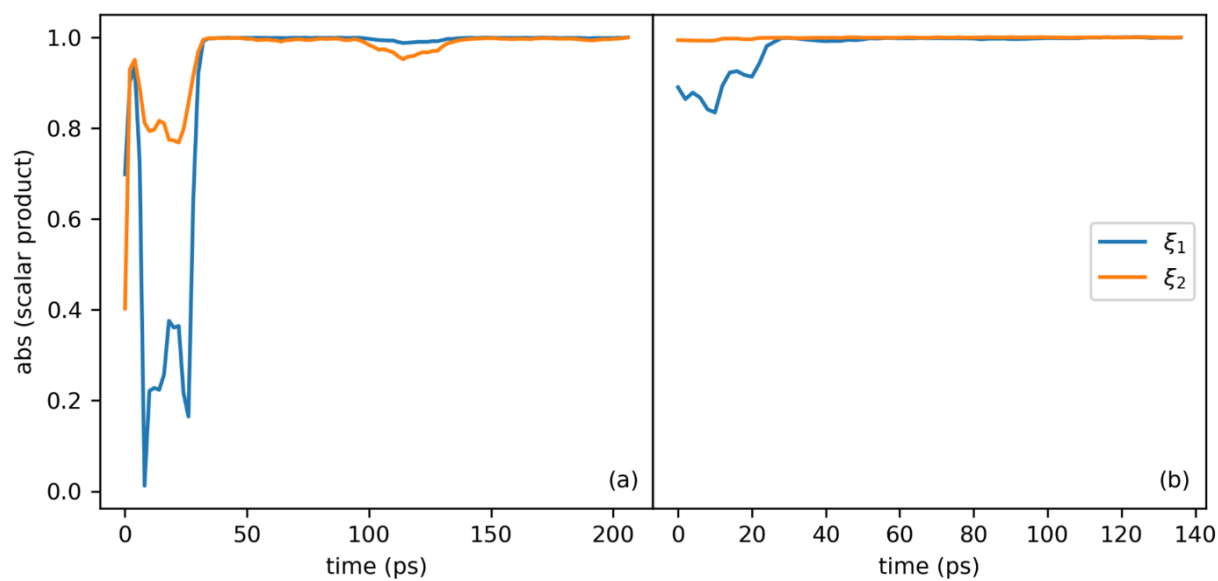

**Figure S1.** *Converged behavior of HLDA CVs, $\xi_1$ and $\xi_2$, (a) for **1a + 1b** reaction and (b) for **2a + 2b**.*

|  |  | $d_1$ | $d_2$ | $d_3$ | $d_4$ | $d_5$ | $d_6$ | $d_7$ |
|---|---|---|---|---|---|---|---|---|
| **1a + 1b** | $\xi_1$ | - 0.353 | - 0.388 | 0.581 | 0.573 | - 0.223 | - 0.059 | - 0.078 |
|  | $\xi_2$ | - 0.579 | - 0.582 | 0.295 | 0.239 | 0.399 | 0.109 | 0.108 |
| **2a + 2b** | $\xi_1$ | 0.327 | 0.340 | 0.280 | 0.319 | - 0.717 | - 0.193 | - 0.215 |
|  | $\xi_2$ | - 0.491 | - 0.498 | 0.514 | 0.494 | - 0.040 | - 0.028 | - 0.026 |

**Table S1**. *$\xi_1$ and $\xi_2$ are linear combinations of $d_{1-7}$, in this table their coefficients are reported.*

Because of the discrimination properties of HLDA, it is already possible to have an insight into the reaction looking at the differences of the signs in Table S1. $\xi_1$ and $\xi_2$ identify a projection in $d_{1-7}$ descriptor subspace, so their global sign does not have a *per-se* meaning. On the other hand, a particular CV well describes a process that implies the elongation of all the distances with coefficients of a given sign and the contraction of all the distances with coefficients of the opposite sign. Thus, this CV is a good identifier between the starting and final states of this process. For example for **2a + 2b**, $\xi_1$ distinguishes well between reactants and products, while $\xi_2$ between the two products. Moreover, from sign accordance it is possible to see that in all the reaction of interest $d_1$ and $d_2$ elongate or contract together, as well as for $d_3$ and $d_4$ couple and for $d_5$, $d_6$ and $d_7$.

# S2 Error Assessment and Convergence Control of the Free Energy Differences between Reactants and Products.

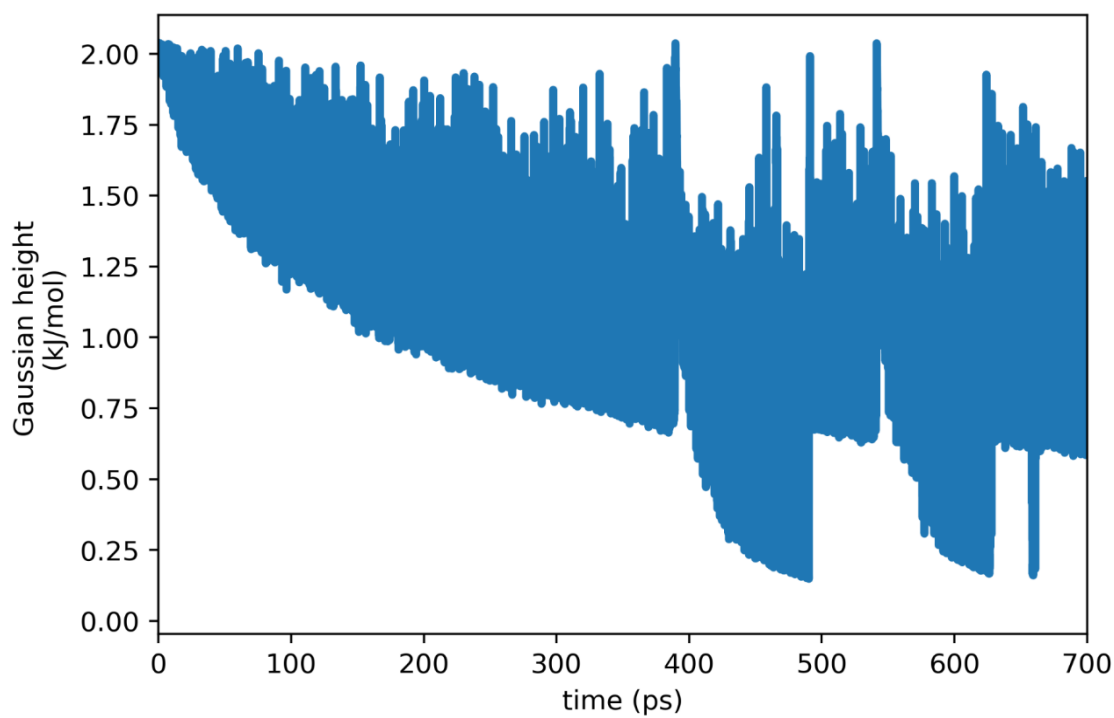

*Figure S2.* Gaussians height as function of time. At the start of the simulation, the system is visiting the reactants basin and the height is exponentially decreased. When it moves towards the first product (at $t=380$ ps), the height of the Gaussians reaches again the starting value of 2, and then it decreases faster, being this area narrower that the reactants one. When it goes back into reactants basins ($t=500$ ps), height turns again to the value it had at $t=380$ and keeps on decreasing. As in the previous case, visiting the second product (from $t=550$ ps), height first assumes values of 2 and then rapidly decrease.

|  | 1a + 1b | 2a + 2b |
|---|---|---|
| $\tau$ | $5 \cdot 10^{-2}$ | $5 \cdot 10^{-2}$ |
| $h_0$ | 0.02 | 0.01 |
| $\gamma$ | 55 | 60 |
| $\sigma_0$ | $9.6 \cdot 10^{-2}$ | $2.6 \cdot 10^{-1}$ |
| $t$ | $5 \cdot 10^3$ | $5 \cdot 10^3$ |

*Table S2.* Parameter for the adaptive-variance WT-MetaD simulations. $\tau$ is the Gaussian deposition rate and $t$ the simulation time (ps). $h_0$ is the initial Gaussian height and $\sigma_0$ its initial standard deviation (kJ/mol). $\gamma$ is the bias factor.

We controlled the convergence of the WT-metaD runs monitoring in time the free energy ($F$) differences between the basins (Figure S3), computed according to:

$$F_{basin} = -k_B T \ln\left( \int_{basin} d\xi_1 d\xi_2 \; e^{-\frac{F(\xi_1,\xi_2)}{k_B T}} \right), \qquad (S1)$$

with $k_B$ the Boltzmann constant, $T$ the temperature. After t=7 ns, it starts fluctuating around a converged value.

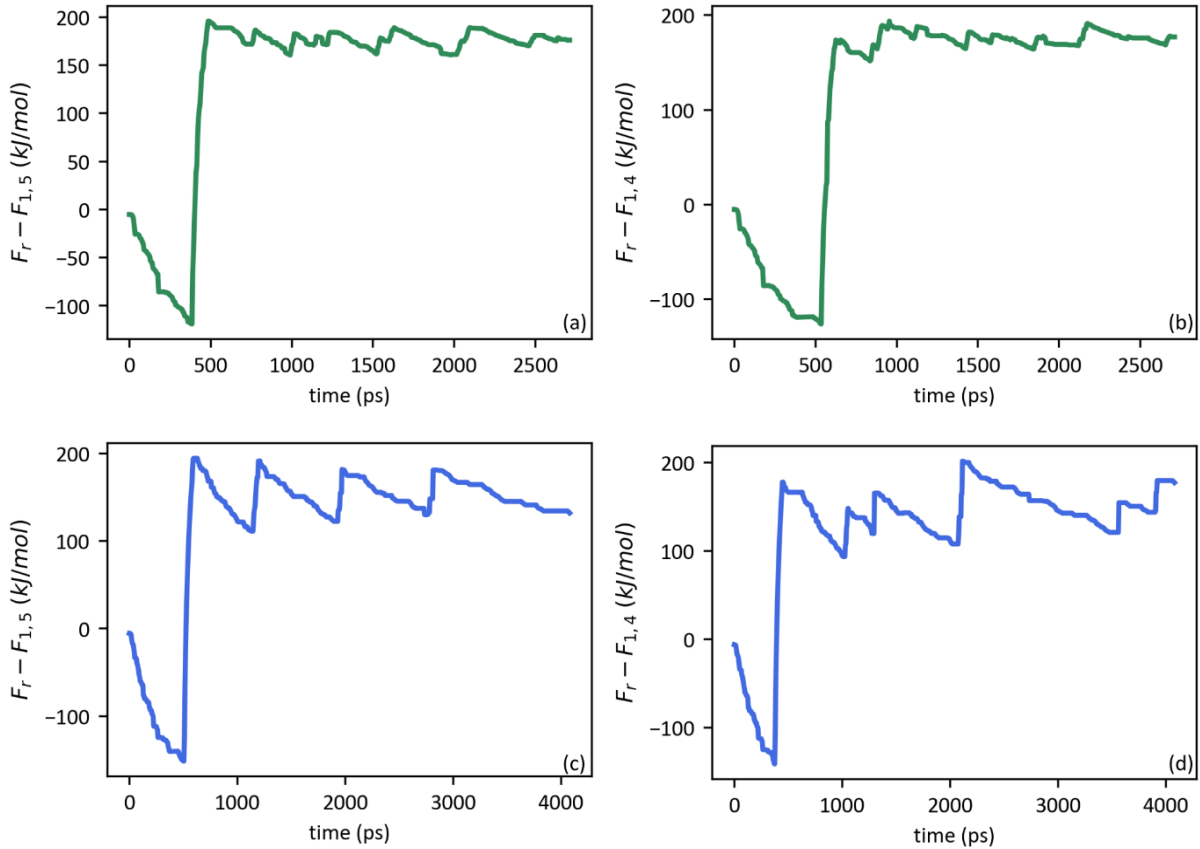

*Figure S3. Differences in free energy of the basins in time. Upper panels: **1a + 1b** reaction, lower panel: **2a + 2b** reaction*

*The final estimation of the free energy differences is associated to a statistical error, using block average on the converged data (from 0.7 ns to 3 ns).*[1]

## S3 Transition State Ensembles Characterization

In order to discriminate between the ensembles of Figure 5 of the manuscript, an additional CV is needed, which we name $\xi_3$. For 1,4 triazole $\xi_3=d_1-d_2$, while for 1,5 triazole $\xi_3=d_3-d_4$. To get the unbiased probability the FES (function of $\xi_1$ and $\xi_2$) was reweighted on $\xi_1$, $\xi_2$ and $\xi_3$ and the associate free energy was calculated using Eq. (S1). Finally, the probability was obtained according to:

$$p = e^{-F/k_B T}.$$

## S4 Restraint Geometry Optimization for 2a + 2b

Likewise the **1a + 1b** case the results are commented again in light of the restraint geometry optimizations (Figure S4). The resulting PES preserve the features of the **1a + 1b** ones. As a consequence, the geometries found in $t_{1,4}$ and $t_{1,5}$ regions show strong analogies to **1a + 1b** system, with the exception of $t_{1,5}^c$ case, that is not found for **2a + 2b** system. This absence can be explained if we consider that there is an additional effect of increased steric hindrance due of the increased size of the substituents. The latter is responsible also for the inversion in probability of the concerted ($t_{1,4}^c$) and the stepwise states ($t_{1,4}^{s1}$) towards 1,4 triazole.

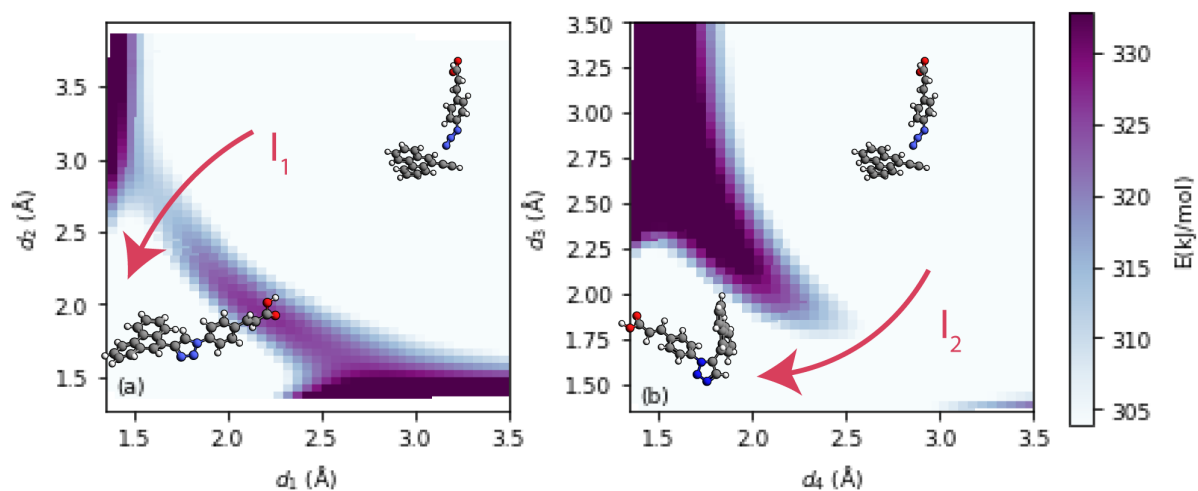

*Figure S4. Energy grid obtained with restraint geometry optimizations for **2a + 2b** reaction.*